# Weighted Naïve Bayes Model for Semi-Structured Document Categorization

*Pierre-François Marteau*[a], *Gildas Ménie*[a], *Eugen Popovic*[a]
[a] *VALORIA Laboratory, University of South-Britany, Campus de Tohannic, 56000 Vannes, France*

The aim of this paper is the supervised classification of semi-structured data. A formal model based on bayesian classification is developed while addressing the integration of the document structure into classification tasks. We define what we call the structural context of occurrence for unstructured data, and we derive a recursive formulation in which parameters are used to weight the contribution of structural element relatively to the others. A simplified version of this formal model is implemented to carry out textual documents classification experiments. First results show, for a adhoc weighting strategy, that the structural context of word occurrences has a significant impact on classification results comparing to the performance of a simple multinomial naïve Bayes classifier. The proposed SCANB implementation competes on the Reuters-21578 data with the SVM classifier associated or not with the splitting of structural components. These results encourage exploring the learning of acceptable weighting strategies for this model, in particular boosting strategies.

Keywords: I Bayesian Classification, Semi-Structured Text Categorization, XML, Weighting Fusion Heuristics.

## 1 Introduction

Since 1990, text classification has become a very active area (cf. [17] for a review) in the Information Retrieval community. A non exhaustive list of approaches based on machine learning includes, Naïve Bayes decision [6], [8], [11], [12], [16], k-nearest neighbours [9], Support Vector Machine (SVM) [7], [10], [21]. Decision Trees and decision rules [1] and some Meta approaches such as boosting [18] and stacking [19], [22]. More recently, growing volume of semi-structured documents available either on the World Wide Web or within traditional data bases increases drastically the needs for specific information retrieval algorithms able to cope with the composite nature of these documents.

This paper contributes to the attempt to exploit the structural knowledge when categorizing semi-structured text documents. From a naïve bayes classification perspective and from the definition of XML structural context of word occurrences, we first develop a formal model. We then concentrate on a simplified implementation of this model to address experimentation on the Reuters database [15]. The results obtained are compared with other approaches such as naïve bayes classification on flat text or SVM based classification with or without the integration of structural knowledge of semi-structured documents.

## 2 XML Context Modeling

A semi-structured document $d$ is well represented by a tree structure $T_d$, containing a set of vertices $S_d$ and a set of edges $A_d$. Within the DOM tree structure associated to a well formed XML document, each node $n$ (in particular each leaf) is connected to the root of the tree by mean of a unique path that we will refer as $c(n)$. Following earlier developments initiated in the field of approximate searching in semi-structured XML data [13], this path is an ordered sequence of XML elements that determines the occurring context of node $n$ inside the document. In case where a leaf $l$ of the DOM tree can be decomposed into a set of sub-elements $\{\nu_i\}$ we will consider that each sub-element $\nu_i$ is attached to the XML context $c(l)$. In particular, if a leaf is identified as a textual element, each word $\nu_i$ (lemma, stem, string, etc.) is attached to the XML context $c(l)$.

More precisely, $c(n)$ is identifiable to an ordered sequence of XML elements attached to the nodes of the path connecting node $n$ to the root of the DOM tree:

$c(n) = <e(n_0), a(n_0)><e(n_1), a(n_1)> \ldots <e(n_p), a(n_p)>$ with :

- $n_0$ the root node,
- $n_p$ is the ancestor node of $n_{p-1}$,
- $p+1$ is the length (number of nodes) of the path $c(n)$.
- $e(n_i)$ is the XML element attached to node $n_i$ (« TITLE », « BODY », « PCDATA », etc.)
- $a(n_i)$ is the set of <attribute, value> pairs potentially attached to node $n_i$.

For a decomposable leaf $l$, each sub-element $\nu_i$ belonging to the decomposition of $l$ is considered as a terminal node without attribute. In such case, the occurring context of sub-element $\nu_i$ will be:

$$c(\nu_i) = <e(n_0), a(n_0)><e(n_1), a(n_1)> \ldots <e(n_p), a(n_p)><e(l), a(l)>.$$

```
<?xml version="1.0" encoding="ISO-8859-1"?> <FILE>
 <REUTERSTOPICS="NO" LEWISSPLIT="TEST" CGISPLIT="TRAINING-SET"
     OLDID="24" NEWID="21">
      <DATE>19-OCT-1987 15:27:23.12</DATE>
      <TOPICS><D>earn</D></TOPICS>
      <PLACES><D>usa</D></PLACES>
      <PEOPLE/><ORGS/><EXCHANGES/><COMPANIES/>
    <UNKNOWN>5;5;5;F22;22;1;f283231;reuter BC-LANE-TELECOMMUNICATION 10-19 0080 </UNKNOWN>
    <TEXT> 2;
        <TITLE>LANE_TELECOMMUNICATIONS PRESIDENT RESIGNS</TITLE>
        <DATELINE>HOUSTON, Oct 19 </DATELINE>
        <BODY> Lane Telecommunications Inc said Richard Lane, its president and chief operating
officer, resigned effective Oct 23. Lane founded the company in 1976 and has been its president
since its inception, …;
        </BODY>
       </TEXT>
 </REUTERS>
</FILE>
```

**Fig. 1.** XML file Example, extracted from the Reuters corpus ([15]).

We can distinguish several cases for the decomposition of leaf *l*, e.g:

– *l* is identified as a set of sub-element $\{ v_i \}$, in this case we will write: e(*l*)=$\{v_i\}$.
– *l* identified as a sequence of sub-elements $v_1 v_{2\ldots} v_k$, in this case we will write: e(*l*)= $v_1 v_{2\ldots} v_k$ .

An example of XML document is given in Fig. 1 while the corresponding DOM tree is given in Fig. 2. From now on to the rest of the paper, we will focus on the first case. Based on this definition of the structural context for node occurrences we develop hereinafter a model of Structural Context Augmented Naïve Bayesian Classification (SCANB).

## 3 SCANB Model

In the context of a bayesian classification of documents, *a posteriori* probabilities to choose a class $\omega$ given a test document *d* are related to the conditional probabilities $P(d|\omega)$ according to the Bayes's law:

$$P(\omega \mid d) = \frac{P(d \mid \omega) * P(\omega)}{P(d)} \quad (1)$$

If we accept the tree representation (see Fig. 2.) for semi-structured documents, we are lead to assimilate P($d|\omega$) to a conditional probability $P(T_d|\omega)$ where $T_d$ is the tree associated to document *d*. We have to consider two difficulties at this level:

– some assumptions need to be formulated to decompose and simplify the estimation of these probabilities,
– given the heterogeneity of semi-structured documents, the number of parameters to cope with increases rapidly with the size of the covered application domain. The training task could require a large volume of data.

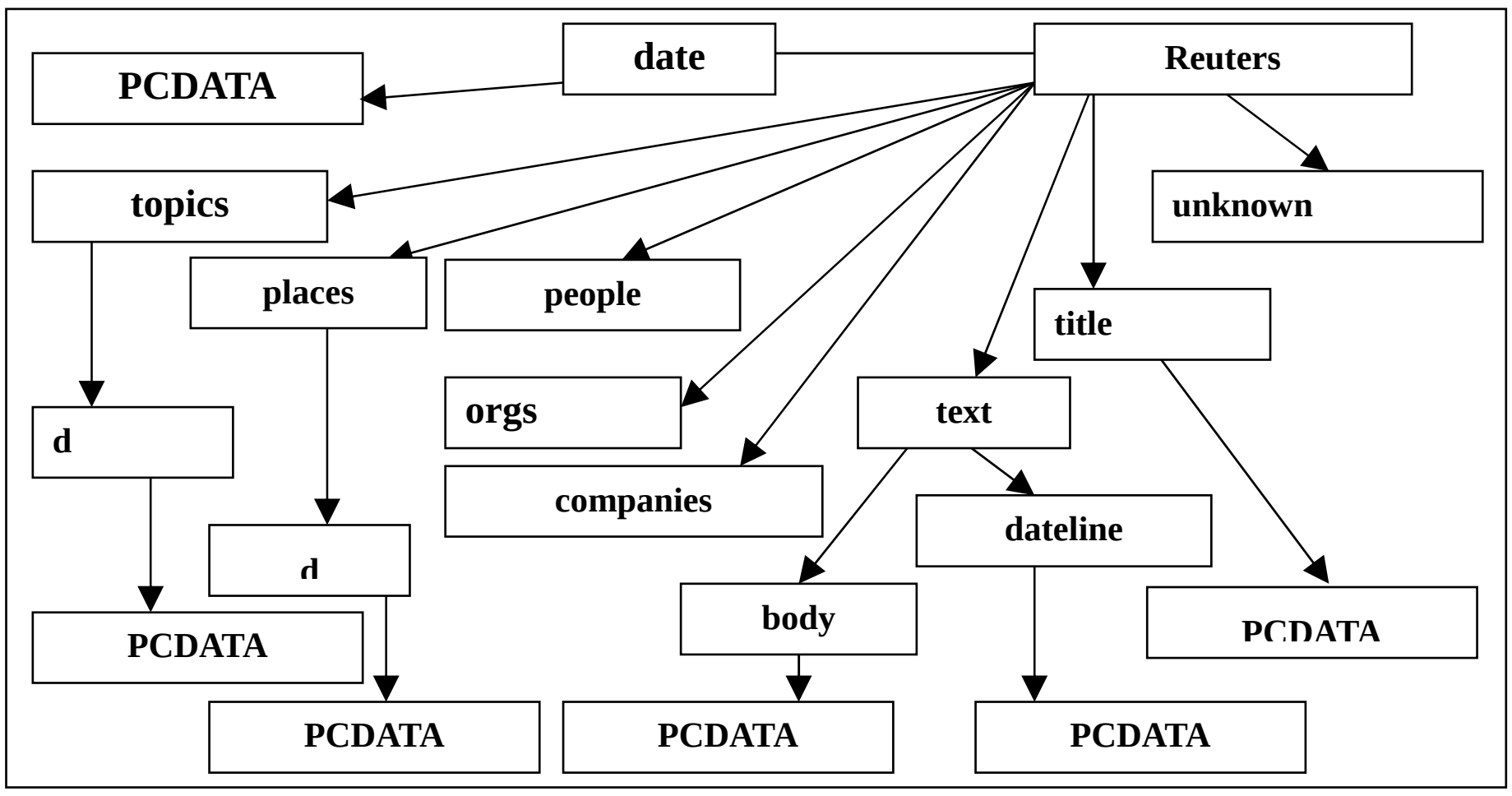

**Fig. 2.** DOM tree for the file presented in Figure 1. The PCDATA elements correspond to leaves (textual nodes).

To address these difficulties that relate to the complexity of the data, some simplifying approaches have been proposed while integrating document structure and content into classification paradigm. Among them, we find: the « splitting » approach [2], [3] which consists in exploiting as many classifier as there are different structural elements (or nodes) into the document data base. The final classification decision is taken according to a fusion function that combines the predictions of the classifiers associated to the structural elements. The Structured Vector Model proposed by Yi and Sundaresan [23] or the Structured Multimedia Document Classifier (SMDC) developed by Denoyer et al. [4] are examples of approaches that somehow generalize the splitting principle. The Structured Vector Model is a data structure dedicated to the representation of trees. The authors have proposed a probabilistic model that integrates local frequencies of terms that depend upon the precise localisation of the text content inside the document structure. When a large number of learning data is available, some experimentation tasks shows that the error rate is significantly lowered comparatively to a classical vector model classifier applied on the flat text. The SMDC is mainly based on two simplifying hypothesis that allow to ease the estimation of the parameters of the model, by relaxing the dependence assumption between the variables: these hypotheses states that the information contents attached to the nodes of the tree structure are independent from each other given the document structure, and furthermore, the information content depends solely on the node to which it is attached. The authors show that on a classification experimentation carried on web HTML data [14], the SMDC model improves significantly the performance of a naive bayes classifier performing on flat text.

The proposed SCANB model lies in between these previous approaches and relates also to some splitting principle. To decompose the probability $P(T_d|\omega)$ we consider the two following hypotheses:

(H1):
$$\begin{aligned} P(T_d / \omega) &= P(r, T_1, T_2, ..., T_k / \omega) \\ &= P(T_1, T_2, ..., T_k / r\,\omega).P(r/\omega) \end{aligned}, \qquad (2)$$

where $r$ is the root node of tree $T_d$ and $\{T_i\}$ is the set of sibling sub-trees accessible from $r$. To simplify the writing, $r$ stands for the node itself or the $\langle e(r), a(r)\rangle$ pair composed with the XML element content $e(r)$ and the set of attributes $a(r)$ attached to the node $r$. This first hypothesis states that the order of occurrence of the sub-trees has no importance.

(H2):
$$\begin{aligned} P(T_d / \omega) &= K_{w,r}.P(T_1 / r\,\omega)^{w_1}.P(T_2 / r\,\omega)^{w_2} ... P(T_k / r\,\omega)^{w_k}.P(r/\omega) \\ &= K_{w,r}.P(r/\omega).\prod_{i=1}^{k} P(T_i / r\,\omega)^{w_i} \end{aligned} \qquad (3)$$

$\{w_i\}$ is a set of positive weighting factors that allow to balance the importance of the relative contribution of each sub-tree. $K_w$ is a normalizing factor. The weightings act as a geometric mean on the sibling nodes. Note that for $w_i=1$ for all $i$, this hypothesis states the conditional independence of the sibling sub-trees given the root node $r$ and the category $\omega$. Given these two assumptions, we can recursively decompose the probability $P(T_d|\omega)$ since $P(T_i|r\ \omega)$ is decomposable into:

$$K_{w,r_i}.P(r_i / r\,\omega).\prod_j P(T_{i,j} / r\,r_i\omega)^{w_j} \qquad (4)$$

where $r_i$ is the root of sub-tree $T_i$, $T_{i,j}$ are sub-trees accessible from $r_i$ and $\{w_j\}$, $K_{w,r_i}$ are the weightings and the normalizing constant attached to node $r_i$. Note that, by the end of the recursion, the weightings for node $r_i$ will integrate the result of the multiplication of the weights positioned along the path from the root to node $r_i$.

As $r_i$ is a sibling of the root $r$ for tree $T_d$, $r$ is identifiable to the XML context (path) of node $r_i$ as previously defined, i.e. $c(r_i)$. In a similar way, '$r\ r_i$' is the XML context of the root node of sub-tree $T_{i,j}$. Thus:

$$K_{w,r_i}.P(r_i / r\,\omega).\prod_j P(T_{i,j} / r\,r_i\omega)^{w_j} = K_{w,r_i} P(r_i / c(r_i)\,\omega).\prod_j P(T_{i,j} / c(r_j)\,\omega)^{w_j} \qquad (5)$$

By the end of the recursion, we obtain the simple formulation:

$$P(T_d / \omega) = \prod_{n_i \in S_d} K_{w,n_i}.P(n_i / c(n_i)\,\omega)^{w_{n_i}} \qquad (6)$$

Formula (6) states that, given hypotheses (H1) and (H2), the estimation of the probability for tree $T_d$ (associated to semi-structured document $d$) conditionally to category $\omega$ is reduced to the weighted products evaluated on the set of nodes $S_d$ of the occurring probability for node $n_i$ given the category $\omega$ and the occurring context of node $n_i$ inside tree $T_d$. To simplify the writing without loss of generality, we consider that weights $w_{ni}$ integrates the weights product associated to $n_i$'s ancestor nodes. The following example in Fig. 3 leads to the decomposition given in formula (7), in which we have only consider a weighting for node 'REUTERS' that possesses two sibling nodes. For all other nodes, the weights are set to unity.

```
<?xml version="1.0" encoding="ISO-8859-1"?>
<FILE>
    <REUTERS ID="21">
      <TITLE>
         texte1…
      </TITLE>
      <BODY>
         texte2…
      </BODY>
    </REUTERS>
</FILE>
```

**Fig. 3.** Simplified example of a Reuters-21578 document.

The example in Fig. 3 leads to the decomposition given in formula (7), in which we have only considered a weighting for node 'REUTERS' that possesses two sibling nodes. For all other nodes, the weights are set to unity.

$$\begin{aligned}
&P(T_d / \omega) = P(< FILE, \varnothing > \mid \omega). \\
&\quad K_{w,reuters}.P(< \text{REUTERS}, \{\text{ID} = "21"\} > \mid < FILE, \varnothing >, \omega). \\
&\quad P(< TITLE, \varnothing > \mid < FILE, \varnothing >< \text{REUTERS}, \{\text{ID} = "21"\} >, \omega)^{w_1}. \\
&\quad P(<"texte1...", \varnothing > \mid < FILE, \varnothing >< \text{REUTERS}, \{\text{ID} = "21"\} >< TITLE, \varnothing >, \omega)^{w_1}. \\
&\quad P(< BODY, \varnothing > \mid < FILE, \varnothing >< \text{REUTERS}, \{\text{ID} = "21"\} >, \omega)^{w_2}. \\
&\quad P(<"texte2...", \varnothing > \mid < FILE, \varnothing >< \text{REUTERS}, \{\text{ID} = "21"\} >< BODY, \varnothing >, \omega)^{w_2}
\end{aligned} \qquad (7)$$

In formula (6), $P(n_i / c(n_i)\omega)$ may characterize complex dependence relationships that must be furthermore detailed. For instance one can use a markov chain to limit the dependence path (for an order *one* dependence, sibling nodes depend only on the root node). This could drastically lower the number of required parameters. Moreover, considering a decomposable leaf *l* of tree $T_d$, we address here the case where the decomposition of *l* is represented by a set $\{ \nu_i \}$ of sub-elements, i.e: $e(l) = \{ \nu_i \}$ thus:

$$\begin{aligned}
P(l / c(l)\,\omega) &= P(< e(l), a(l) > / c(l)\,\omega) = P(e(l) / a(l)\,c(l)\,\omega).P(a(l) / c(l)\,\omega) \\
&= P(\{\nu_i\} / a(l)\,c(l)\,\omega).P(a(l) / c(nl)\,\omega)
\end{aligned} \qquad (8)$$

The last hypothesis states the conditional independence of characteristics $\nu_I$ :

$$\text{(H3)} \qquad P(l / c(l)\,\omega) = P(a(l) / c(l)\,\omega) \prod_i P(\nu_i / a(l)\,c(l)\,\omega) \; . \qquad (9)$$

## 5 Parameter Estimation for Text Categorization

From this point to the end of the paper, we will not consider attributes attached to node and will focus on textual elements. The SCANB model relies on the three hypotheses (H1), (H2) and (H3) and is defined by the set of equations: $\omega^* = \arg\max_{\omega \in \Omega} \{P(\omega / T_d)\} = \arg\max_{\omega \in \Omega} \{P(T_d / \omega).P(\omega)\}$

$$P(T_d / \omega) = \prod_{n_i \in S_d} K_{w,n_i}.P(n_i / c(n_i)\,\omega)^{w_{n_i}} \qquad (10)$$

$$P(l / c(l)\,\omega) = \prod_i P(\nu_i / c(l)\,\omega)$$

with: $T_d$ the tree representing the semi-structured document *d*, $n_i$ is a node in tree $T_d$, *l* is a leaf that decomposes into a set of independent characteristics $\nu_i$, $c(n_i)$ is the path from node $n_i$ to the root of tree $T_d$, $w_{ni}$ is the weights for node $n_i$.

The estimation of probabilities *P(l/c(l)w)* is handled according to the following equation:

$$\prod_i P(v_i / c(l)\,\omega) = \frac{N^{l,d}!}{\prod_i N_i^{l,d}!} \prod_i \left(\theta_i^{l,\omega}\right)^{N_i^{l,d}} , \qquad (11)$$

where $N_i^{l,d}$ is the number of time characteristic $v_i$ occurs in the XML element attached to node $l$ of tree $T_d$, $N^{l,d} = \sum_i N_j^{l,d}$ and $\theta_i^{l,\omega}$ the probability that characteristic $v_i$ occurs in node $l$ for class $\omega$.

We evaluate $\theta_i^{l,\omega}$ using Laplace's estimator: $\theta_i^{l,\omega} = \frac{N_i^{l,\omega} + 1}{N^{l,\omega} + |V_l|}$ (12)

Where $N_i^{l,\omega}$ is the number of time characteristic $v_i$ occurs in the set of XML elements attached to node $l$ of training documents for class $\omega$,

$N^{l,\omega} = \sum_i N_j^{l,\omega}$ and $|V_l|$ the size of the vocabulary for node $l$.

In practice, we will use the following decision function: $\omega^* = \arg\max_{\omega \in \Omega} \{Log(P(T_d / \omega).P(\omega))\}$, i.e.:

$$\omega^* = \arg\max_{\omega \in \Omega} \left\{ \begin{array}{l} Log(P(\omega)) + LK_w + \\ \sum_{n_i \in S_d} w_{n_i}.Log(P(n_i / c(n_i)\,\omega)) \end{array} \right\}, \qquad (13).$$

where $LK_w = \sum_{n_i} Log(K_{w,n_i})$.

This formulation shows that the decision function is nothing but a linear combination (whose coefficients are the weights $w_{ni}$) of the elementary decision rules attached to each node elements in $S_d$. Various learning policies or adhoc strategies can be proposed to set up these weight parameters.

# 6 Experimentation

## 6.1 Experimental Dataset

The Reuters-21578 [25] database is widely exploited for the validation and comparison of text categorization models. In our experiments, we selected the ten most frequent categories from this corpus as our dataset for training and testing. The articles are distributed unevenly across the 10 categories with the largest category containing 3964 documents and the smallest only 286: *Ω = {Acq, Corn, Crude, Earn, Interest, Ship, Trade, Grain, Money-fx, Wheat}.*

## 6.2 Pre-processing

We simplify the XML structure of the documents by retaining within the $T_d$ tree the paths connecting the textual leaves to the root. Considering the Reuters data, we have retain *<TITLE>*, *<BODY>* and *<DATELINE>* as textual elements, we segment the elements text value into words according to the following separator characters: « `.;,:?<>=+}{()'\"^$#[]\\/\n` ». Reuters-21578 is a *multilabeled* dataset, so we train a binary *one-vs-rest* classifier for each category. This choice to retain the previous three XML elements and to reject the others has been governed by the concern to compare our results with other studies.

## 6.3 Weight Heuristic for the SCANB Model

To balance the node contributions according to a vocabulary coverage principle, we propose the following heuristic for the weight parameters of the SCANB model:

$$w_n = \left\{ \begin{array}{l} \frac{|V_{n,\omega}|}{|d_n|}, \; if\; n \,has\; a\, textual\, element\, attached \\ 1,\; otherwise \end{array} \right\},$$

where $|V_{n,\omega}|$ is the cardinal of the vocabulary associated to node $n$ for the category ω and $|d_n|$ stands for the size of the textual element attached to node $n$ inside document $d$. As the weights $w_n$ are independent from the category, the $LK_w$ term is the same for all categories and the decision rule can reduce to:

$$\omega^* = \arg\max_{\omega\in\Omega}\left\{ Log(P(\omega)) + \sum_{n\in S_d\cap T} \frac{|V_{n,\omega}|}{|d_n|}.Log(P(n/c(n)\omega)) \right\}, \quad (14)$$

where $T=\{<TITLE>, <BODY>, <DATELINE>\}$.

The aim of this heuristic is, for a given document *d*, to increase the contribution of a node *n* for which the size of the attached textual element is high comparatively to the vocabulary size associated to *n*, and conversely, to penalize the contribution of a node *n* for which the size of the attached textual element is low comparatively to the vocabulary size associated to *n*. Thus, the weight associated to node *n* in document *d* is a kind of quality criteria that measures the vocabulary coverage of node *n* for document *d*.

### 6.4 Measures Used for Evaluation

To carry on comparative experimental studies, we will use the $F_1$ measure [20] defined as the harmonic mean of two complementary measures: precision (*P*) and recall (*R*):

$$F_1(P,R) = \frac{2\cdot P}{P+R}, \text{ where } P = \frac{TP}{TP+FP} \text{ and } R = \frac{TP}{TP+FN}. \quad (15)$$

Precision and recall are two standard measures widely used in text categorization literature to evaluate an algorithm's effectiveness [20] on a given category $\omega$.In the formulas above, for a category $\omega$, the true positives (*TP*) is the number of documents belonging to category $\omega$ that are correctly classified as category $\omega$; the false positives (*FP*) is the number of documents not belonging to category $\omega$ that are incorrectly classified as category $\omega$; the false negative (*FN*) is the number of documents belonging to category $\omega$ that are incorrectly classified as non-category $\omega$ by a classifier. For the multilabeled Reuters-21578 dataset, we report on micro-averaged recall, (*R*) precision (*P*), and $F_1$ measures calculated as follows:

$$P = \frac{\sum_{\omega_i\in\Omega} TP(\omega_i)}{\sum_{\omega_i\in\Omega} TP(\omega_i) + \sum_{\omega_i\in\Omega} FP(\omega_i)}, \quad R = \frac{\sum_{\omega_i\in\Omega} TP(\omega_i)}{\sum_{\omega_i\in\Omega} TP(\omega_i) + \sum_{\omega_i\in\Omega} FN(\omega_i)} \quad (16)$$

In our experiments, we use the ten-fold cross validation method to evaluate the previous measures and to compare the classification models on the Reuters dataset.

### 6.5 Comparative Study with SVM Classifier as Reported by Bratko and Filipic

The above procedure has been applied for the 10 categories:*Ω = {Acq, Corn, Crude, Earn, Grain, Interest, money-fx, Ship, Trade, Wheat}*, using <TITLE>, <DATELINE> and <BODY> XML components as specified in the work reported by Bratko and Filipic [2][3]. We can thus compare the Naive Bayes on flat text (NB), the Naïve Bayes with Splitting (NBS), the SCANB model with the weightings given above, the SVM on flat text (SVM) and the SVM with splitting (SVMS) models. Our results on SCANB are given in table 2, while Bratko and Filipic's results are given in table 1. In our Experiments, a NBS' model close to the NBS model has been derived and evaluated from the SCANB model by setting all the weights equal to unity.

**Table 1.** Precision, recall and $F_1$ measure for models NB,NBS, SVM and SVMS from Bratko and Filipic work on Reuters-21578 database

| *Measure* | *NB* | *NBS* | *SVM* | *SVMS* |
|---|---|---|---|---|
| *Recall* | 0.9623 | 0.9548 | 0.9214 | 0.9660 |
| *Precision* | 0.8280 | 0.8485 | 0.9658 | 0.9178 |
| $F_1$ | 0.8901 | 0.8985 | **0.9431** | 0.9413 |

**Table 2.** Precision, recall and $F_1$ measure for models NB, NBS and SCANB models on Reuters-21578 database, according to our tests

| Measure | NB | NBS | SCANB |
|---|---|---|---|
| *Recall* | 0.9185 | 0.9241 | 0.9540 |
| *Precision* | 0.8540 | 0.8586 | 0.9294 |
| $F_1$ | 0.8851 | 0.8902 | **0.9415** |

On these two experiments, we notice that the NBS models perform slightly better than the NB model as previously shown by other studies [4][23]. This is corroborated by Bratko and Filipic [2][3] that find similar results when comparing the naive bayes classifier applied on flat text or in conjunction with a splitting method. On the other hand the SCANB model, with the proposed weightings heuristic shows significant improvements since a 5.6% gain is achieved on the $F_1$ measure against the naïve bayes model applied on flat text and a 5.1% gain is achieved on the $F_1$ measure against the NBS' model. This experimentation shows that a proper weighting strategy, in the context of the Reuters-21578 data base, has a significant impact on the classification accuracy. The proposed weightings strategy performs quite comparatively to the SVM model that seems insensitive to the structural organization of documents.

## 7 CONCLUSION

We have considered the problem of supervised categorization for semi-structured data. A formal model named SCANB in between approaches proposed by Yi and Sundaresan [23], Denoyer an al. [4] or the splitting procedure described in Bratko and Filipic [2][3] has been developed. This model extends traditional naïve Bayesian classification while integrating document structure knowledge. Given some acceptable hypotheses, we get a recursive development for the model that is original to our knowledge. It allows approximating the tree structure of semi-structured data as a set of nodes from which the path to the root is attached. For each node, a weighting parameter is added to balance the importance of the node comparatively to the others according to some heuristics. First results performed on the reuters-21578 database shows that the structure of documents plays a crucial role as it improve the classification efficiency. Given an ad hoc heuristic independent from the categories, the SCANB model attain the same level of accuracy as the SVM model applied on flat text or after the splitting of textual components. The linear form of the decision function given in equation 13 suggests that a boosting approach [18] [24] can be easily experimented.